\documentstyle[psfig,aps,prl,amssymb,multicol]{revtex}
\begin{document}
\draft
\title{Extremal collision sequences of particles on a line:
\\ optimal transmission of kinetic energy} 
\author{Thorsten P\"{o}schel$^{1}$ and Nikolai V.~Brilliantov$^{1,2}$}

\address{$^1$Humboldt-Universit\"at zu Berlin, Charit\'e, Institut f\"ur Biochemie,
 Monbijoustra{\ss}e 2. D-10117 Berlin, Germany }
\address{$^2$Moscow State University, Physics Department,
   Moscow 119899, Russia} 
\date{\today}
\maketitle
\begin{abstract}
The transmission of kinetic energy through chains of inelastically
colliding spheres is investigated for the case of constant
coefficient of restitution $\epsilon={\rm const}$ and impact-velocity
dependent coefficient $\epsilon(v)$ for viscoelastic
particles. We derive a theory for the optimal distribution of particle
masses which maximize the energy transfer along the chain and check it 
numerically. We found  that for $\epsilon={\rm const}$ the mass
distribution is a monotonous function which does not depend on the
value of $\epsilon$. In contrast, for $\epsilon(v)$ the mass distribution 
reveals a pronounced maximum, depending on the particle properties
and on the chain length. 
The system investigated demonstrates that
even for small and simple systems the velocity dependence of the
coefficient of restitution may lead to new effects with respect to the 
same systems under the simplifying approximation $\epsilon={\rm const}$.

\end{abstract}


\begin{multicols}{2}

\section{Introduction}
\label{sec:intro}
Chains of nonlinear interacting particles have been of large interest
since a long time and a variety of interesting effects occurring in
those systems has been described, such as solitons,
(e.g.~\cite{solitons}), energy localization (e.g.~\cite{localization}), 
etc.  In the context of granular materials chains of
inelastically colliding particles have been investigated as model
systems for shaken granular material (e.g.~\cite{Luding,TPTS}),
granular compaction~\cite{carparking}, the ``inelastic collapse'' 
(e.g.~\cite{Shida,DuLiKadanoff}). The kinetic theory of one-dimensional 
granular systems has been addressed in \cite{oneDkinetics}.

In this paper we consider a linear chain of inelastically colliding
particles of masses $m_i$, radii $R_i$ ($i=0\dots n$) with initial
velocities $v_0=v >0$ and $v_i=0$ ($i=1\dots n$) at initial positions
$x_i>x_j$ for $i>j$ with $x_{i+1}-x_i> R_{i+1}+R_i$ (Fig.~\ref{fig:sketch}). The masses of the
first and last particles $m_0$ and $m_n$ are given and we address
the questions: How have the masses in between to be chosen to maximize
the energy transfer from the first particle of the chain to the last one?
If $n$ is variable, how should $n$ be chosen to maximize the 
after-collisional velocity $v_n^{\prime}$ of the last particle. 

\begin{minipage}{8cm}
  \begin{figure}[htbp]
    \centerline{\psfig{figure=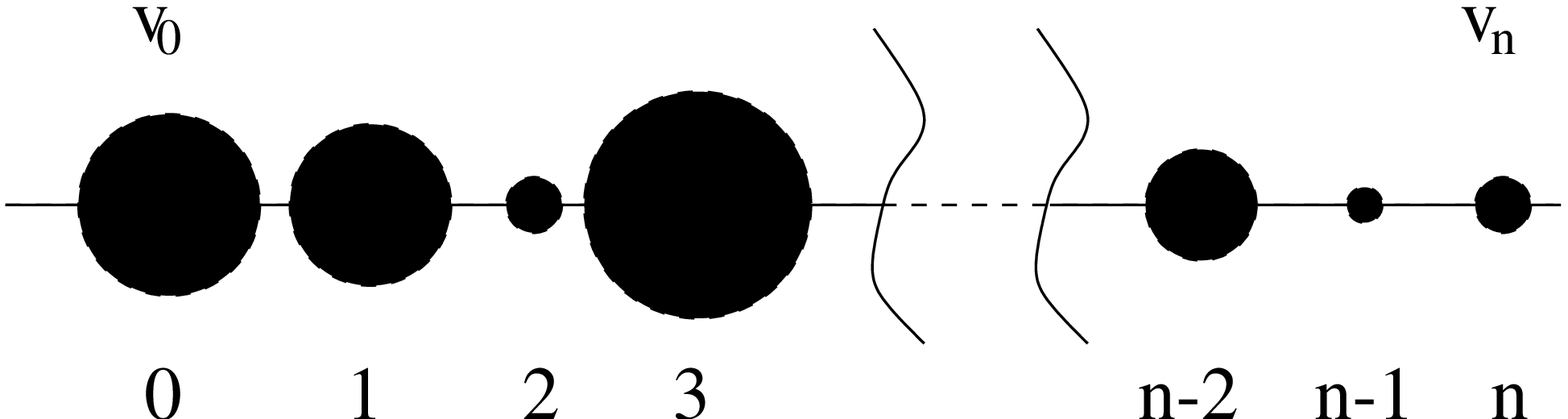,width=8cm}}
    \caption{Sketch}
\vspace{0.5cm}
    \label{fig:sketch}
  \end{figure}
\end{minipage}

One can easily study the chains of ideally elastic spheres and of spheres 
interacting  via a constant coefficient of restitution. It is much more 
complicated to deal with chains of of viscoelastic particles, 
which have an impact velocity dependent coefficient and which, as we show 
below, exhibit quite unexpected behavior. It has been demonstrated recently 
that the kinetic
properties of ``thermodynamically-large'' systems of viscoelastic particles differ
significantly from those of particles interacting with constant
coefficient of restitution~\cite{NBTP}. The system considered in this
paper may serve as an example of a {\em small} system which properties
change qualitatively when the viscoelastic properties of the particles
are taken into account explicitly.

In the present study the problem of the most efficient energy
transmission in a chain of particles of variable mass is addressed. 
We analyze the optimal distribution for the particle masses  and 
calculate the optimal size of the system. 

\section{Elastic Particles}

The textbook problem of elastic collisions may serve us to introduce
the notation. Assume particle 0 collides with the resting particle
1. Then after the collision the velocity of particle 1 is
\begin{equation}
\label{1}
v_1^\prime=\frac{2m_0}{m_0+m_1} \, v_0
\end{equation}
(the primed variables 
refer to after-collisional velocities) and for a chain of $n+1$ particles of 
masses $m_0$, $m_1$ \dots $m_n$ one has analogously \cite{remarkmult}
\begin{equation}
\label{4}
v_{n}^\prime= 2^n \prod_{k=0}^{n-1} \left(1+\frac{m_{k+1}}{m_k}
\right)^{-1}\, v_0\,.
\end{equation}
For this system one finds easily that the choice
$m_i=\sqrt{m_{i-1}m_{i+1}}$, $(i=2\dots n-1$) maximizes $v_n^{\prime}$.  If we
fix $m_0$ and $m_n$, obviously the mass distribution
\begin{equation}
\label{5}
m_k=\left(\frac{m_n}{m_0}\right)^{k/n}m_0 
\end{equation}
maximizes $v_n^\prime$:
\begin{equation}
v_n^\prime=\left[ \frac{2}{1+\left(\frac{m_n}{m_0}\right)^{1/n}}
\right]^{n} \,v_0
\end{equation}
The function $R_v=v_n^\prime/v_0$ always increases with $n$ and has
the limit
\begin{equation}
\label{7}
R_v=\left[ \frac{v_n^{\prime}}{v_0} \right]_{n \to \infty} = 
\sqrt{\frac{m_0}{m_n}}\,,
\end{equation}
i.e., if the masses of the particles are chosen according to (\ref{5})
the kinetic energy of the first particle is completely transferred to
the last one by a chain of infinite length.

For the case of dissipative collisions an infinite chain cannot be
optimal since in each collision energy is dissipated. Hence, we expect
an optimum for the chain length for which the velocity of the last
particle reaches its maximum.

\section{Particles with a constant restitution coefficient}

According to our model the particles collide pairwise. This 
allows to use the restitution coefficient, which relates the 
relative velocity of colliding particles $i$ and $i+1$ after 
collision to that before the collision:

\begin{equation}
\label{epsdef00}
\epsilon = 
\left|\frac{v_{i+1}^{\prime}-v_i^{\prime} }{v_{i+1}-v_i} \right|\,.
\end{equation}
Equation (\ref{1}) turns then into
\begin{equation}
\label{8}
v_1^\prime=\frac{1+\epsilon}{1+\frac{m_1}{m_0}} \, v_0, 
\end{equation}
where we again assume that the particle with velocity $v_0$ and mass
$m_0$ hits a particle of mass $m_1$ at rest, which starts moving with
the velocity $v_1^\prime$. Straightforward generalization of the
previous analysis for the case of the dissipative collisions with  
a constant coefficient of restitution $\epsilon$ shows that the optimal mass 
distribution is identical to that for the elastic case (\ref{5}).
This means that the optimal mass distribution does not
depend on the dissipation if $\epsilon={\rm const}$. The velocity of the 
last particle in the chain reads for this case:

\begin{equation}
\label{6}
v_n^\prime=\left[
\frac{1+\epsilon}{1+\left(\frac{m_n}{m_0}\right)^{1/n}} \right]^{n}
\,v_0\,.
\end{equation}
Figure \ref{fig:constantMass} shows the
optimal mass distribution for different chain lengths $n$. The mass of
the first particle is $m_0=1$ and of the last particle $m_n=0.1$.

In the next section we will consider particles which interact via a
velocity dependent coefficient of restitution. Since the velocity of
the particles varies for the particles of the chain we characterize
the dissipation of the colliding spheres not by the coefficient of
restitution itself but rather we define a dissipative constant
$b$. For the case of a constant coefficient $\epsilon$ it is defined
as $b=(1-\epsilon)$. 

\begin{minipage}{8cm}
  \begin{figure}[htbp]
    \centerline{\psfig{figure=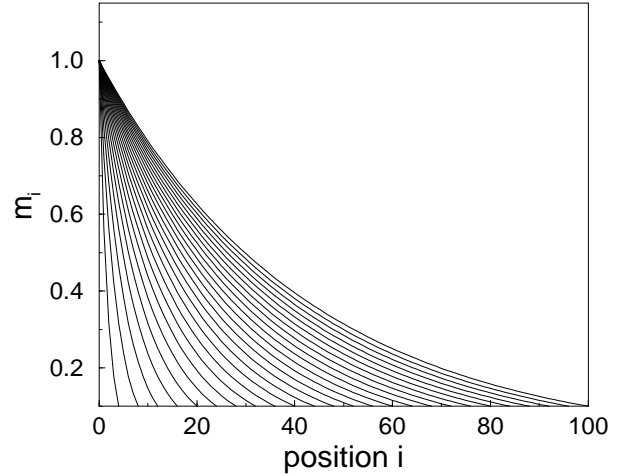,width=8cm,angle=0}}
\vspace{0.3cm}
    \caption{Optimal mass distribution $m_i$, $i=1\dots n$,  for the case of a 
constant restitution coefficient $\epsilon$. Each of the lines shows the mass $m_i$ over the index $i$ for a specified chain length $n$. The masses of the first and last particles are fixed at $m_0=1$ and $m_n=0.1$.}
\vspace{0.5cm}
    \label{fig:constantMass}
  \end{figure}
\end{minipage}

\begin{minipage}{8cm}
  \begin{figure}[htbp]
    \centerline{\psfig{figure=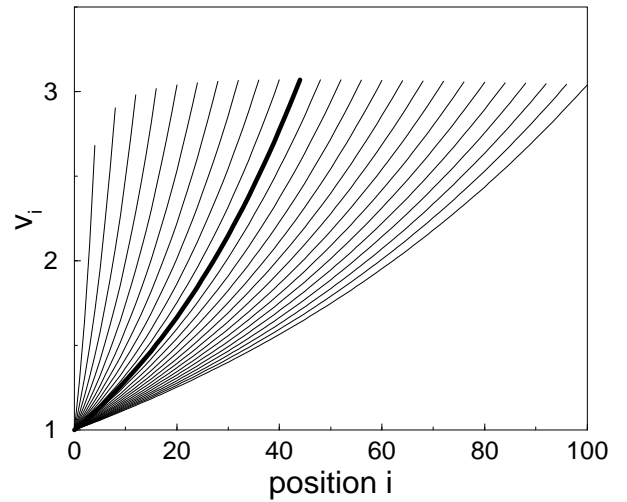,width=8cm,angle=0}}
\vspace{0.3cm}
    \caption{Velocity distribution of particles in chains 
    with the optimal mass distribution (given in Fig.~\ref{fig:constantMass}) 
    according to (\ref{5}). Each of the lines shows the velocity 
    $v_i$ over the index $i$ for a specified chain length $n$.
    The dissipative constant is
    $b=(1-\epsilon)=5\cdot 10^{-4}$. 
    The last particle reaches its maximal velocity
    for chain length $n^*=44$ (bold drawn). The velocity of the first
    particle of the chain is $v_0=1$.}
\vspace{0.5cm}
    \label{fig:constmv22v}
  \end{figure}
\end{minipage}

In contrast to the mass distribution the corresponding velocity
distributions do depend on the value of the restitution coefficient
$\epsilon$. Figures \ref{fig:constmv22v} and \ref{fig:constmv42v} show
the velocity distribution for two different values of the dissipative
constant, $b=5\cdot 10^{-4}$ and $b=0.032$.

\begin{minipage}{8cm}
  \begin{figure}[htbp]
    \centerline{\psfig{figure=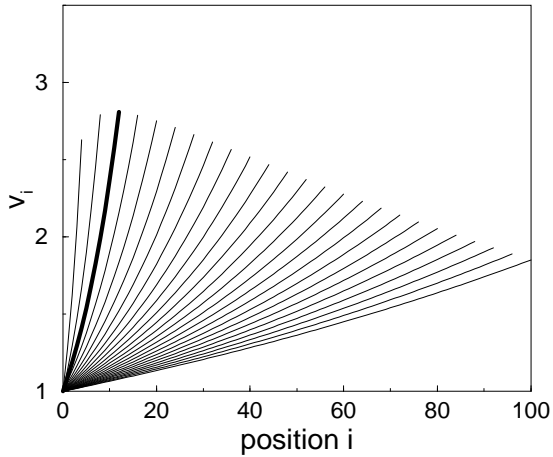,width=7.3cm,angle=0}}
\vspace{0.1cm}
    \caption{Same as Fig.~\ref{fig:constmv22v} but for $b=0.032$. The
    optimal chain length is $n^*=12$}
\vspace{0.5cm}
    \label{fig:constmv42v}
  \end{figure}
\end{minipage}
\begin{minipage}{8cm}
  \begin{figure}[htbp]
    \centerline{\psfig{figure=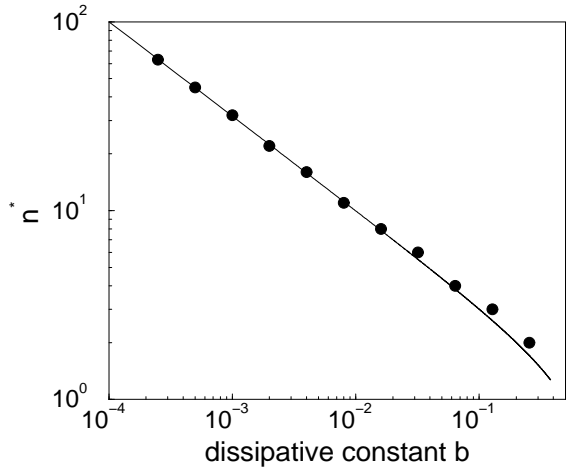,width=7.3cm,angle=0}}
\vspace{0.1cm}
    \caption{The optimal chain length $n^*$, which gives the 
    maximal transmission 
    of energy along the chain with the fixed first and last masses, as a 
    function of the dissipative
    parameter $b =(1-\epsilon)$. The line shows the prediction of
    Eq. (\ref{optnconst}), with $x_0$ found numerically. Points refer
    to the results of a direct numerical optimization of the masses in
    the chain.}
\vspace{0.5cm}
    \label{fig:noptconst}
  \end{figure}
\end{minipage}

For the case of dissipative collisions the ratio $R_v=v_n^\prime/v_0$
does not monotonously increase with $n$, but rather it has an extremum
which shifts to smaller chain lengths with increasing dissipative
parameter $b$. The optimal value of $n$, which maximizes $R_v$ reads:
\begin{equation}
\label{10}
n^*=\frac{\log \left( m_n/m_0 \right)}{\log\left( x_0 \right) }
\label{optnconst}
\end{equation}
where $x_0$ is the solution of the equation
\begin{equation}
\label{11}
(1+x_0)=(1+\epsilon)x_0^{x_0/(1+x_0)}
\end{equation}
Correspondingly, the extremal value of the $R_v$ reads
\begin{equation}
\label{12}
R_v^*=\left[ \frac{1+\epsilon}{1+x_0} \right]^{n^*}
\end{equation}
In Fig.~\ref{fig:noptconst} the dependence of the extremal $n^*$ on the
restitution coefficient is shown.

\section{Viscoelastic Particles}
\subsection{Collisional law for the viscoelastic particles}

It has been shown that for colliding viscoelastic spheres the
restitution coefficient depends on the masses of the colliding particles
and also on their relative velocity $v_{ij}$ \cite{BSHP}. An explicit
expression for the coefficient of restitution is given by the series
\cite{TomThor,Rosa}
\begin{equation}
\epsilon=1-C_1\left(\frac{3A}{2} \right)\alpha^{2/5} v_{ij}^{1/5}\!+
\!C_2 \left(\frac{3A}{2} \right)^2\alpha^{4/5} v_{ij} ^{2/5}\! \mp\! \cdots
\label{epsilon}
\end{equation}
with
\begin{equation}
\alpha= \frac{2~ Y\sqrt{R^{\,\mbox{\footnotesize\em eff}}}}{
3~ m^{\mbox{\footnotesize\em eff}}\left( 1-\nu ^2\right) }
\label{rhodef}
\end{equation}
where $Y$ is the Young modulus and $\nu$ is the Poisson ratio. The
effective mass and effective radius is defined as
$R^{\,\mbox{\footnotesize\em eff}}=R_iR_j/(R_i+R_j)$,
$m^{\mbox{\footnotesize\em eff}}=m_im_j/(m_i+m_j)$ where $R_{i/j}$ and
$m_{i/j}$ are radii and masses of the colliding particles. The
constant $A$ describing the dissipative properties of the spheres
depends on material parameters (for details see~\cite{BSHP}).  The
constants $C_1=1.15344$ and $ C_2=0.79826 $ were obtained analytically
in Ref.~\cite{TomThor} and then confirmed by numerical simulations.

For the following calculation we neglect terms ${\cal
O}\left(v^{2/5}\right)$ and of higher-orders. Moreover we also assume
for simplicity that all particles are of the same radius $R$, but have
different masses \cite{bdef}. We abbreviate
\begin{equation}
\label{epsviab}
\epsilon=1-b\frac{v_{ij}^{1/5}}{\left(m^{\,{\mbox{\footnotesize\em eff}}} 
\right)^{2/5}}
\end{equation}
with
\begin{equation}
b=C_1 \left(\frac{3A}{2} \right)\left(\frac{2}{3}\frac{Y
\sqrt{R/2}}{1-\nu^2}\right)^\frac25.
\label{bdef}
\end{equation}
Thus, the collision with $\epsilon = {\rm const}$ and 
given dissipative constant $b$, as introduced above,  corresponds 
(i.e. has equal value of $\epsilon$) to the viscoelastic collision with 
the same $b$, with unit effective mass $m^{\,{\mbox{\footnotesize\em eff}}}=1$ 
and unit relative velocity $v_{ij}$.

Hence, for viscoelastic particles the velocity of the $k+1$-rst
particle after colliding with the $k$-th reads
\begin{equation}
\label{vkk}
v_{k+1}^{\prime} =
\frac{2-b\left(\frac{m_{k+1}+m_k}{m_{k+1}m_k} \right)^{2/5} 
v_k^{1/5} }{1+\frac{m_{k+1}}{m_k}} \, v_k\,.
\end{equation}
The masses $m_k$, $k=1\dots n-1$ which maximize $v_n^\prime$ can be
determined numerically and the results are shown in
Figs. \ref{fig:mv22m} and \ref{fig:mv42m} for two different values of
the dissipative constant $b$.

\begin{minipage}{8cm}
  \begin{figure}[htbp]
    \centerline{\psfig{figure=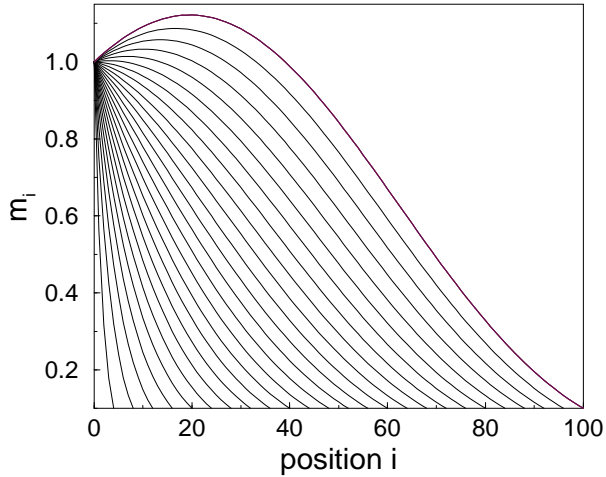,width=8cm,angle=0}}
\vspace{0.3cm}
    \caption{Optimal mass distribution $m_i$, $i=1\dots n$,  for the case of  
      viscoelastic particles with the restitution coefficient given by 
      (\ref{epsviab}) with $b=5\cdot 10^{-4}$. Each of the lines shows 
      the mass $m_i$ over the index $i$ for a specified chain length $n$. 
      The masses of the first and last particles are  
      $m_0=1$ and $m_n=0.1$.}
\vspace{0.5cm}
    \label{fig:mv22m}
  \end{figure}
\end{minipage}

\begin{minipage}{8cm}
  \begin{figure}[htbp]
    \centerline{\psfig{figure=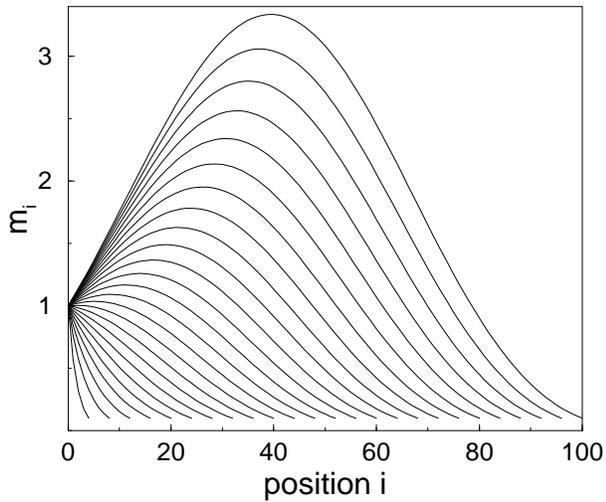,width=8cm,angle=0}}
\vspace{0.3cm}
    \caption{The same plot as Fig.~\ref{fig:mv22m} but for $b=2\cdot
    10^{-3}$.}
\vspace{0.5cm}
    \label{fig:mv42m}
  \end{figure}
\end{minipage}

For small chain length or small $b$, respectively, the optimal mass
distribution are very close to that for the elastic chain as shown in
Fig.~\ref{fig:constantMass}. Again we find a monotonously decaying
function for the masses. For larger chain length $n$ or larger
dissipation $b$, however, the mass distribution is a non-monotonous
function. The according velocities of the particles in chains of
spheres of optimal masses are drawn in Figs.~\ref{fig:mv22v} and
\ref{fig:mv42v}. Note that the mass distribution and velocity 
distribution are related by Eq. (\ref{vkk}).

\begin{minipage}{8cm}
  \begin{figure}[htbp]
    \centerline{\psfig{figure=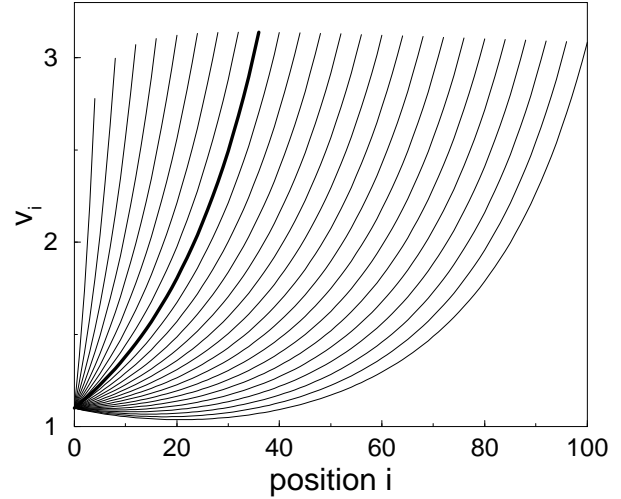,width=8cm,angle=0}}
\vspace{0.3cm}
    \caption{Velocity distribution for viscoelastic particles 
    in chains with the optimal mass distribution given in 
    Fig.~\ref{fig:mv22m}. Each of the lines shows the velocity 
    $v_i$ over the index $i$ for a specified chain length $n$.
    The dissipative constant is $b=5\cdot 10^{-4}$.
    The last particle reaches its maximal velocity
    for the chain length $n^*=36$ (bold drawn). The velocity of the first
    particle of the chain is $v_0=1$.}
\vspace{0.5cm}
    \label{fig:mv22v}
  \end{figure}
\end{minipage}

\begin{minipage}{8cm}
  \begin{figure}[htbp]
    \centerline{\psfig{figure=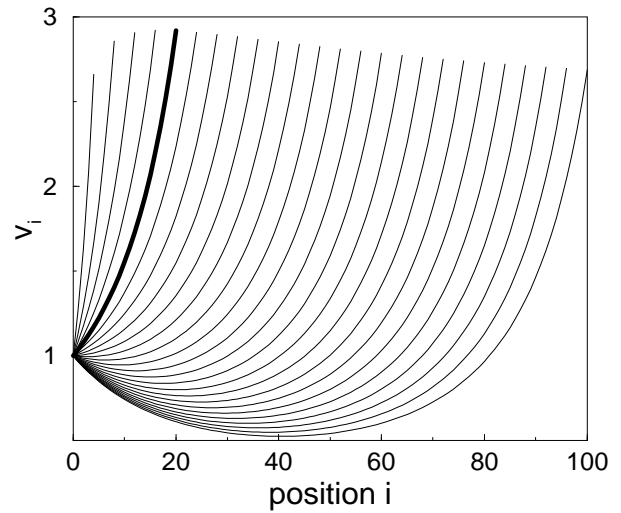,width=8cm,angle=0}}
\vspace{0.3cm}
    \caption{The same plot as Fig.~\ref{fig:mv22v} but for the 
      mass distribution according to Fig.~\ref{fig:mv42m} 
    ($b=2\cdot 10^{-3}$). The optimal chain length is $n^*=20$.}
\vspace{0.5cm}
    \label{fig:mv42v}
  \end{figure}
\end{minipage}

\subsection{Variational approach to the optimal mass-distribution}

In the following we describe an approximative theory of the optimal
collision chain of viscoelastic particles. To this end we first
evaluate the loss of kinetic energy in the chain which we divide
into two parts and term as ``inertial'' and ``viscous'' losses. In our
approach we treat the part of energy which is not transformed from the
first particle of the chain to the last one as a ``lost'' energy.  In
this sense the energy is ``lost'' according to two mechanisms: First,
due to mismatch of subsequent masses, which causes incomplete transfer
of momentum even for elastic collision when the masses differ (this part of
the energy loss is called ``inertial''). The second refers to the
dissipative nature of collisions and, therefore, this loss is called
``viscous'' below. The inertial loss in the collision, attributed to
the energy transfer to the $i$-th particle is thus given by the
energy which remains in the $i-1$-rst particle after the collision:
\begin{eqnarray}
\label{Ein1}
\Delta E_{in}^{(i)} &=& \frac{m_{i-1}}{2}
\left(v_{i-1}^\prime\right)^2\nonumber\\
& =& \frac{m_{i-1}}{2}\left(\frac{m_i-m_{i-1}}{m_i+m_{i-1}}\right)^2
v_{i-1}^2
\end{eqnarray}
For long enough chains we approximate the discrete mass distribution
by a continuous one, $m(x)$. This, with the assumption of small mass
gradients, gives $m_i\approx m_{i-1}+\frac{d m(x)}{d x}\cdot 1 $,
where we assume that particles are separated on a line by a unit
distance. Within the continuum picture $\Delta E_{in}^{(i)} \to
\frac{d E_{in}}{d x} \cdot 1$, and  we write for the ``line-density'' of the
inertial loss, discarding high-order mass gradients:
\begin{equation}
\label{Ein2}
\frac{d E_{in}}{d x} \approx \frac{\left(\frac{dm(x)}{dx}\right)^2}
{8m(x)}v(x)^2
\end{equation}

Viscous losses describe the energy losses according to the inelastic
properties of the material, therefore, they are equal to the
difference of the kinetic energy of a particle after an {\em elastic}
collision (with no dissipation) and that after a {\em dissipative}
collision.

\begin{eqnarray}
&&  \Delta E_{vis}^{(i)} = \left.\frac{m_{i} v_{i}^{2}}{2}
\right|_{\epsilon=1}-\left.\frac{m_{i} v_{i}^2}{2}\right|_{\epsilon = 
\epsilon\left(v_i\right)}= \\
&& \frac{m_{i}}{2}\left(\frac{2}{1+\frac{m_{i}}{m_{i-1}}}\right)^2 
v_{i-1}^2 -  \frac{m_{i}}{2}\left(\frac{1+\epsilon\left(v_{i-1}\right)}
{1+\frac{m_{i}}{m_{i-1}}}\right)^2 v_{i-1}^2 = \nonumber \\
&&\frac{2 m_{i} v_{i-1}^2}{\left( 1+\frac{m_{i}}{m_{i-1}} 
\right)^2} 
\left\{1-\left[1-\frac{b}{2}\left(\frac{m_i+m_{i-1}}{m_i m_{i-1}}
\right)^{2/5} v_{i-1}^{1/5}\right]^2 \right\}
\nonumber
\end{eqnarray}
Now we assume that the dissipative parameter $b$ is small, so that one
can keep only the linear term, expanding $\Delta E_{vis}^{(i)}$ with
respect to $b$. Transforming then to continuous variables and
discarding terms which are products of $b$ and mass gradients (which
are also supposed to be small), yields:

\begin{equation}
\frac{d E_{vis}}{d x} \approx \frac{b}{2^{3/5}} m^{3/5} v^{11/5}.
\end{equation}

Thus, the total energy loss in the entire chain reads
\begin{equation}
\label{Etot1}
  E_{tot} = \int\limits_0^n \left[
\frac{m_x^2}{8m}v^2 + \frac{b}{2^{3/5}} m^{3/5} v^{11/5} \right] dx\,.
\end{equation}
where $m_x \equiv \frac{dm}{dx}$. As it follows from
Eq. (\ref{Etot1}), to evaluate $E_{tot}$ one needs the velocity
distribution $v(x)$. As a zero-order approximation we use an ``ideal
chain Ansatz''. This refers to a velocity distribution $v(x)$ in an
idealized chain, where the kinetic energy completely transforms
through the chain, i.e. where $\frac12 m(x)v^2(x)= {\rm const}=\frac12
m_0v_0^2$. With $m_0=1$, $v_0=1$, so that $v(x)=1/\sqrt{m(x)}$, this
Ansatz yields:

\begin{equation}
\label{etot}
  E_{tot} = \int\limits_0^n \left[ \frac{m_x^2}{8m^2} + \frac{b}{2^{3/5}} 
\frac{1}{m^{1/2}}\right] dx\,.
\end{equation}
The mass distribution which minimizes $E_{tot}$ satisfies the
Euler-equation applied to the integrand in (\ref{etot}):
\begin{equation}
\label{eq:Euler}
  \frac{d}{dx}\frac{2m_x}{8m^2} - 
\frac{\partial}{\partial m} \left[ \frac{m_x^2}{8m^2} + \frac{b}{2^{3/5}}
 \frac{1}{m^{1/2}} \right] = 0\,.
\end{equation}
Eq. (\ref{eq:Euler}) leads to an equation for the mass distribution of
the optimal chain, written for $y(x) \equiv 1/m(x)$:
\begin{equation}
\label{eqy}
  \frac{d^2 y}{dx^2} - \frac{1}{y}\left(\frac{dy}{dx}\right)^2 -
  2^{2/5}b y^{3/2} = 0
\end{equation}
Multiplying Eq. (\ref{eqy}) by $2\left(y^{\prime}/y^2 \right)$ 
($y$ is always positive) we recast (\ref{eqy}) into the form:
\begin{equation}
\label{eqy1}
\frac{d}{dx} \left[ \left( y^{\prime}/y \right)^2 - 
4 \cdot 2^{2/5}b y^{1/2} \right]=0
\end{equation}
which implies the first integral of this equation:

\begin{equation}
\label{eqy2}
 \left( y^{\prime}/y \right)^2 - 
4 \cdot 2^{2/5}b y^{1/2} = -c \, ,  
\end{equation}
where the constant $c$ depends on parameter $b$, the chain length $n$ and 
initial and final masses, $m_0$ and $m_n$. The form of the solution depends 
on the sign of this constant. If the mass distribution has an extremum at 
$x=x^*$, such that  $m^{\prime}(x^*)=0$  and   $y^{\prime}(x^*)=0$, the constant 
$c$ is positive. This follows from Eq. (\ref{eqy2}), i.e., 
$c=4 \cdot 2^{2/5}b y^{1/2}(x^*) > 0$, since  
$y^{1/2}(x^*)$ is positive. 

The solution of the {\it first-order} equation (\ref{eqy2}) may be found 
straightforwardly. The general solution is somewhat  lengthy, but for the 
case of $m_0=1$ (one can always use the appropriate mass unit), this reads
(for $c>0$):

\begin{equation}
\label{eqy3}
y(x)=m(x)^{-1}=\frac{c^2}{2^{4/5}b^2} \cos^{-4}\left(\frac{x\sqrt{c}}{2}+\varphi \right)
\end{equation}
where
\begin{equation}
\label{eqy4}
\cos \varphi = \sqrt{\frac{c}{2^{2/5}b}} \, .  
\end{equation}
The value of the constant $c$ may be found from the second boundary 
condition $y(n)=1/m_n$, which yields a transcendental equation for $c$:

\begin{equation}
\label{eqy5}
\cos\left(\frac{n\sqrt{c}}{2} \right) -
\sin\left(n\frac{\sqrt{c}}{2} \right)\sqrt{\frac{2^{2/5}b}{c}-1}
=\frac{2^{2/5}b}{c}m_n^{-1/4}
\end{equation}
The last equation has to be solved numerically. Instead, however, 
we solved numerically directly the initial differential Eq. (\ref{eqy}). 

Note, that some scaling properties of the solution may be
deduced just from the form of (\ref{eqy}). Namely, as it follows from this
equation, the solution should depend on the reduced length variable
$x\sqrt{b}$. Thus, the distribution of masses for chains with
different chain length $n$ and different dissipative constant $b$
should coincide after rescaling the particle numbers as $i \to
\sqrt{b} \cdot i$, provided masses $m_0$ and $m_n$ are the same for
these chains. We will consider the scaling properties of the mass
distribution in more detail latter.

Figure \ref{fig:mAnalNumA} shows the optimal mass distribution for a
chain of length $n=40$ for different damping parameters $b$. The lines
display the (numerical) solution of the variation Eq. (\ref{eqy})
whereas the points show the results of a numerical optimization of the
chain problem. For small dissipation $b$ both results agree.

For larger values of $b$ the solution of variational equation
(\ref{eqy}) deviates from the results of the numerical
optimization. This follows from the fact that for larger $b$ the
gradients of the mass distribution are not small and our variational
approach loses its accuracy. Note, however, that while the absolute
values of masses in the mass distribution deviate from that given by
variational approach, this still predicts well the position of the
maximum of the distribution.  Figure~\ref{fig:mAnalNum} shows the same
data as Fig.~\ref{fig:mAnalNumA} but for larger dissipation parameter
$b$.
\begin{minipage}{8cm}
  \begin{figure}[htbp]
    \centerline{\psfig{figure=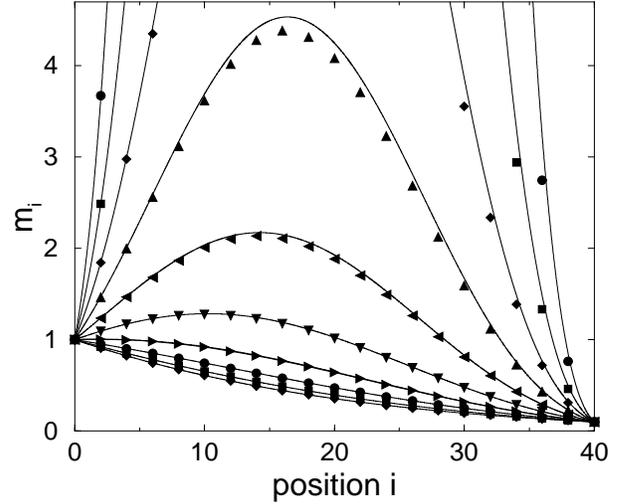,width=8cm,angle=0}}
\vspace{0.3cm}
    \caption{Mass distribution in chains of viscoelastic particles  
      of length $n=40$ with 
      optimal mass distribution for different values of the dissipative 
      parameter $b$. Lines:
    results of the variational theory, according to Eq. (\ref{eqy}), 
    Points: numerical
    optimization (from top to bottom: $\bullet:~b=0.128$,
    $\blacksquare:~b=0.064$, $\blacklozenge:~b=0.032$,
    $\blacktriangle:~ b=0.016$, $\blacktriangleleft :~b=0.008$,
    $\blacktriangledown :~b=0.004$, $\blacktriangleright :~b=0.002, $
    etc.). 
    As previously, $m_i$ is the mass of the $i$-th particle along the chain.}
\vspace{0.5cm}
    \label{fig:mAnalNumA}
  \end{figure}
\end{minipage}

\begin{minipage}{8cm}
  \begin{figure}[htbp]
    \centerline{\psfig{figure=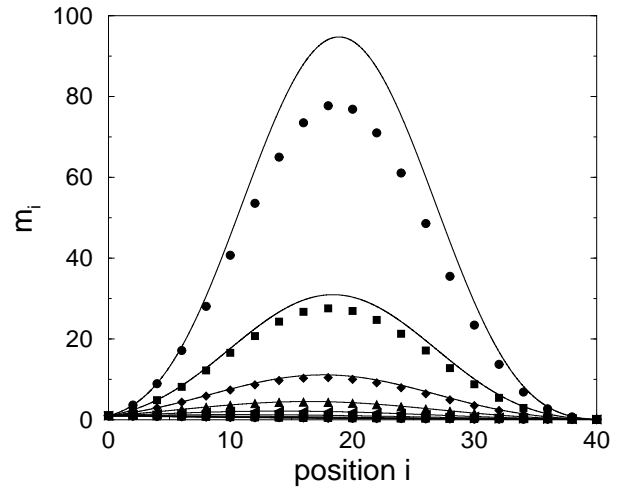,width=8cm,angle=0}}
\vspace{0.3cm}
    \caption{Same data and symbols as in Fig.~\ref{fig:mAnalNumA}  
      but plotted in larger scale.}
\vspace{0.5cm}
    \label{fig:mAnalNum}
  \end{figure}
\end{minipage}

Figure \ref{fig:vepsonv40} displays the velocity distribution 
for the optimal chain with the mass distribution shown 
in Fig.~\ref{fig:mAnalNum}. The data given in Fig.~\ref{fig:vepsonv40}
refer to the numerical optimization where Eq. (\ref{vkk}), which 
relates velocity and mass distribution is used.  
 According to the maximum in the mass
distribution, the velocity distribution reveals for larger $b$ a
pronounced minimum.

One can give a simple physical explanation of appearance of maximum in
mass distribution (and correspondingly minimum in the velocity
distribution): As it is seen from Eq. (\ref{epsviab}) the restitution
coefficient increases with decreasing impact velocity and increasing
masses of colliding particles; this reduces the viscous losses. Thus
slowing down particles, by increasing their masses in the inner part
of the chain, leads to decrease of the viscous losses of the energy
transfer.  The larger the masses in the middle and the smaller their
velocities, the less energy is lost due to dissipation. On the other
hand, since masses $m_0$ and $m_n$ are fixed, very large masses in the
middle of the chain will cause large mass mismatch of the subsequent
masses and thus large inertial losses [see Eq. (\ref{Ein1})].  The
optimal mass distribution, minimizing the {\em total} losses
compromises (dictated by $b$) between these two opposite tendencies.
For the case of a constant coefficient of restitution the relative
part of the kinetic energy, which is lost due to dissipation does not
depend on the impact velocity. This means that only minimization of
the inertial losses, caused by mass gradient, may play role in
optimization of the mass distribution. Thus only a monotonous mass
distribution with minimal mass gradients along the chain may be
observed as an optimal one for the case of the constant restitution
coefficient.

\begin{minipage}{8cm}
  \begin{figure}[htbp]
    \centerline{\psfig{figure=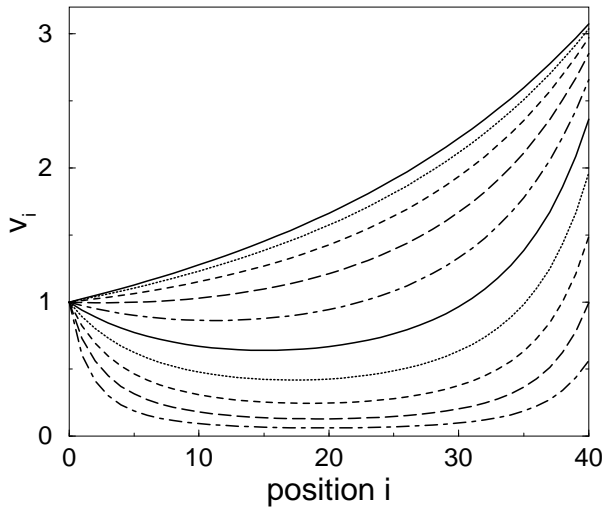,width=8cm,angle=0}}
\vspace{0.3cm}
    \caption{The velocity distribution in chains of 
      viscoelastic particles  of length
    $n=40$ with the optimal mass distribution according to
    Fig.~\ref{fig:mAnalNum} for different values of the dissipative
    constant $b$. Lines from top to bottom: $b=2.5\cdot 10^{-4}$,
    $5\cdot 10^{-4}$, $0.001$, $0.002$, $0.004$, $0.008$, $0.016$,
    $0.032$, $0.064$, $0.128$. 
     The velocity distribution is obtained from the mass distribution
     (given in Fig.~\ref{fig:mAnalNum}) according to Eq. (\ref{vkk}).
     As previously, $v_i$ is the velocity  of the $i$-th 
    particle along the chain.}
\vspace{0.5cm}
    \label{fig:vepsonv40}
  \end{figure}
\end{minipage}

As in the case of the constant restitution coefficient, the velocity
of the last particle $v_n^{\prime}$ of an {\em optimal} chain depends on
$n$. For short chains (with $m_0$, $m_n$ fixed) the mass gradient of
adjacent particles is large, hence inertia losses are large as
well. For very long chains viscous losses become large. Hence, we
expect that among the optimal chains exists a chain with a certain
length $n^*$ which allows for an optimal transmission of kinetic
energy from the first particle to the last
one. Fig.~\ref{fig:vendepsonv} shows the velocity of the last particle
for chains with optimal mass distribution as a function of the chain
length $n$ for different values of the dissipative parameter $b$.
Naturally, as for the case of constant restitution coefficient, the
optimal chain length $n^*$ shifts to smaller values with increasing
dissipative constant $b$.

\begin{minipage}{8cm}
  \begin{figure}[htbp]
    \centerline{\psfig{figure=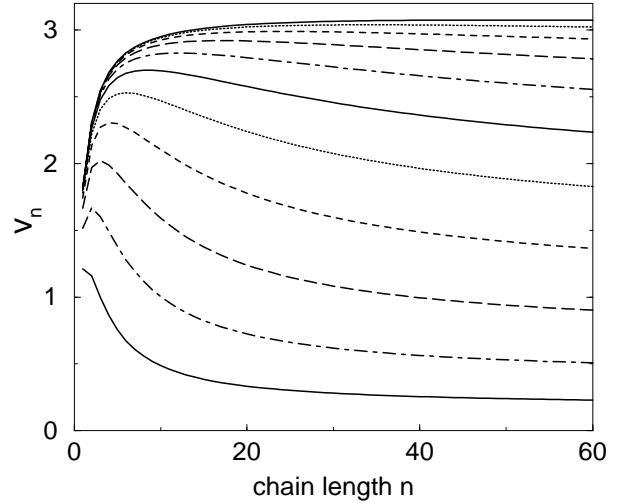,width=8cm,angle=0}}
\vspace{0.3cm}
    \caption{Velocity of the last particle $v_n$ for chains of 
      viscoelastic particles with optimal
    mass distribution over the chain length $n$ for different values 
    of $b$. As in Fig.~\ref{fig:vepsonv40}, the velocity distribution 
    was obtained from the mass distribution according to Eq. (\ref{vkk}), 
    and lines from top to bottom correspond to  $b=2.5\cdot 10^{-4}$,
    $5\cdot 10^{-4}$, $0.001$, $0.002$, $0.004$, $0.008$, $0.016$,
    $0.032$, $0.064$, $0.128$. Note that with increasing 
    dissipative constant $b$ the maximum
    of $v_n(n)$, which corresponds to the optimal chain length $n^*$ 
    shifts to smaller values of $n$, which means naturally, that 
    optimal chains are shorter for larger dissipation. }
\vspace{0.5cm}
    \label{fig:vendepsonv}
  \end{figure}
\end{minipage}

Having the mass distribution and the velocity distribution obtained from 
the numerical optimization one can check directly the validity of the 
"ideal chain Ansatz", $v(x)=1/\sqrt{m(x)}$, used in the variational approach. 
In Fig.~\ref{fig:optimAnsatz} we compare $v(x)$ obtained by 
optimization with that from the Ansatz. As it is seen from the figure, 
the ideal chain Ansatz occurs to be rather accurate for small dissipation 
parameter $b$, and for the initial part of the chain. It demonstrates, 
however, noticeable deviations from the optimization
data for larger $b$, especially at the end of the chain, i.e. for 
$i \approx n$. This is not surprising since it uses an assumption of 
complete transmission of energy, which is definitely poor for the very end 
of the chain. On the other hand, as it follows 
from Figs. \ref{fig:mAnalNumA} and \ref{fig:mAnalNum}, 
this Ansatz yields rather accurate results, when applied 
to the mass distribution problem. The possible explanation for this 
follows from  the boundary condition for the mass distribution 
at the end of the chain, $m(i=n)=m_n$. This  imposes the correct 
behavior of the mass distribution at this part of the chain and partly 
compensates the inaccuracy of the velocity distribution which develops 
mainly at the chain end (see Fig.~\ref{fig:optimAnsatz}).

\begin{minipage}{8cm}
  \begin{figure}[htbp]
    \centerline{\psfig{figure=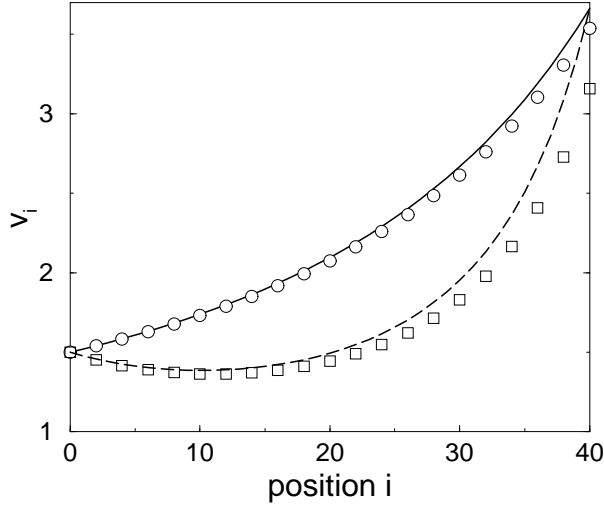,width=8cm,angle=0}}
\vspace{0.3cm}
    \caption{The velocity distribution in chains of 
      viscoelastic particles  of length
    $n=40$ with the optimal mass distribution according to
    Fig.~\ref{fig:mAnalNumA}. Lines give the velocity 
    distribution for the ideal chain Ansatz, $v_i=1/\sqrt{m_i}$ 
    (with masses taken from the optimization data), points show the 
    numerical optimization data for  $b=0.001$ (top) and  
    $b=0.008$ (bottom). Note that for these values of the dissipative
    parameter $b$ the variational theory gives very accurate description 
    for the optimal mass distribution (see Fig.~\ref{fig:mAnalNumA})}
\vspace{0.5cm}
    \label{fig:optimAnsatz}
  \end{figure}
\end{minipage}

\subsection{Scaling laws for  the optimal mass-distribution}

Now we analyze how the maximal mass $m^* \equiv m(x^*)$ (the mass of
the heaviest sphere located at $x=x^*$) in the optimal mass distribution 
depends on the chain
length $n$ and the dissipative parameter $b$. We show that there exists a 
simple scaling relation between these values. 

We start from Eq. (\ref{eqy3}) for the optimal mass distribution 
\begin{equation}
\label{m(x)}
m(x)=\frac{2^{4/5}b^2}{c^2} \cos^4\left(\frac{x\sqrt{c}}{2}+\varphi \right)
\end{equation}
with $c$ and $\varphi$ defined by Eqs. (\ref{eqy4}) and (\ref{eqy5}). The 
condition for the optimal mass
\begin{eqnarray}
\label{m_x(x)}
&& m_x(x^*) =  \\
&&= -\frac{2^{9/5}b^2}{c^{3/2}} \cos^3\left(\frac{x^*\sqrt{c}}{2}+\varphi \right) \sin \left(\frac{x^*\sqrt{c}}{2}+\varphi \right) =0 \, \nonumber 
\end{eqnarray}
implies $\sin \left(\frac{x^*\sqrt{c}}{2}+\varphi \right)=0$ and 
thus the relation between the maximal mass $m^*$ and the constant $c$
\begin{equation}
\label{m*x_and_c}
m^*=\frac{2^{4/5}b^2}{c^2} \cos^4 \left(\frac{x^*\sqrt{c}}{2}+\varphi \right) = \frac{2^{4/5}b^2}{c^2}\,,
\end{equation}
i.e., 
\begin{equation}
\label{c}
c=2^{2/5}b/\sqrt{m^*} \, .
\end{equation}
This allows to write the boundary condition for $m(x)$  at $x=n$:
\begin{equation}
\label{m_n}
m_n=m^* \cos^4\left(\frac{n \sqrt{c}}{2}+\varphi \right)
\end{equation}
or equivalently 
\begin{equation}
\label{m_n1}
\frac{n \sqrt{c}}{2} = 
\arccos \left[ \left( \frac{m_n}{m^*} \right)^{1/4} \right] - \varphi\,.
\end{equation}
Simple analysis shows that $\varphi <0 $ if the optimal distribution 
has a maximum (this follows from the form  of the solution (\ref{m(x)}) and
the requirement that $m(x)$ increases at $x=0$). Thus, one obtains from 
Eqs. (\ref{eqy4}) and (\ref{c}):
\begin{equation}
\label{varphi}
\varphi= -\arccos \left[\left( \frac{m_0}{m^*} \right)^{1/4} \right] \,.
\end{equation}
Using again Eq. (\ref{c}) for the constant $c$ we recast Eq. (\ref{m_n1}) into 
the final form:
\begin{eqnarray}
\label{m_n2}
&&n \sqrt{b} =  \\
&& 2^{4/5} \left( m^* \right)^{1/4} \left\{
\arccos \left[ \left( \frac{m_n}{m^*} \right)^{1/4} \right] 
 +\arccos \left[\left( \frac{m_0}{m^*} \right)^{1/4} \right] \right\}
\nonumber 
\end{eqnarray}
This scaling relation expresses the product $n \sqrt{b}$ in terms of the 
maximal mass $m^*$. For the case of a strongly pronounced maximum in the 
optimal mass distribution, i.e., when 
$m_0/m^* \ll  1$ and $m_n/m^* \ll 1 $, one can expand the $\arccos (x) $ in 
(\ref{m_n2}) to obtain a linear scaling relation between  
$(m^*)^{1/4}$ and $n \sqrt{b}$:
\begin{equation}
\label{m_n3}
n \sqrt{b}=p \, \left( m^* \right)^{1/4} -q \,,
\end{equation}
with 
\begin{eqnarray}
p&=&2^{4/5} \pi\, ,\\
q&=&2^{4/5}\left(m_0^{1/4}+m_n^{1/4} \right)\,. 
\end{eqnarray}

In Fig.~\ref{fig:maxmass} we compare the analytical relation
(\ref{m_n2}) and its linear approximation  (\ref{m_n3}) with the results 
for $m^*$, following
from the numerical optimization for the mass distribution for
different chain lengths and different dissipative constants. As one can
see from Fig.~\ref{fig:maxmass}, the results of the analytical theory
and of the numerical optimization agree well, except for large
dissipation values. We would like to stress that there are no fitting
parameters used.

\begin{minipage}{8cm}
  \begin{figure}[htbp]
    \centerline{\psfig{figure=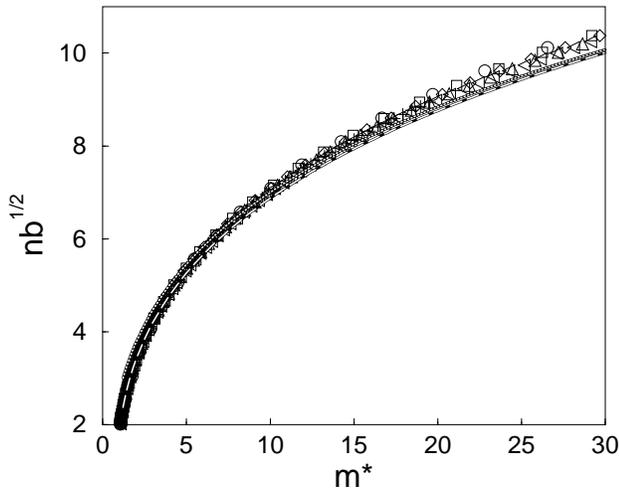,width=8cm,angle=0}}
\vspace{0.3cm}
    \caption{Shows  $n \sqrt{b}$ as a function of $m^*$
      for the chain of viscoelastic particles 
      with the optimal mass distribution. 
      Here $m^*$ is the mass of the heaviest particle in the chain, 
      $n$ is a chain length and  $b$ is the dissipative
      parameter. In the figure we plotted $n\sqrt{b}$ 
      over $m^*$ 
      for about
      3000 different combinations of $b$ and $n$ ($n=2\dots 300$,
      $b=0.0001\dots 0.256$) including all data presented in
      Figs. \ref{fig:mv22m}, \ref{fig:mv42m}, \ref{fig:mAnalNumA} and
      \ref{fig:mAnalNum}. Without any adjustable parameters the data
      from the numerical optimization of chains agrees well with the
      analytical expressions Eq. (\ref{m_n2}), given by the dashed line.
      The linear approximation for the scaling relation, 
      Eq. (\ref{m_n3}) is shown by the dotted line.}
\vspace{0.5cm}
    \label{fig:maxmass}
  \end{figure}
\end{minipage}

Using the optimal mass distribution, Eq. (\ref{m(x)}), one can 
compute the total energy loss in the chain, as given by Eq. (\ref{etot}):
\begin{eqnarray}
\label{E_tot1}
&&E_{tot} = \frac{nc}{2} +2 \sqrt{c}  
\left\{
\frac{ \sqrt{1- \cos^2 \varphi_n}}{\cos \varphi_n} -
\frac{ \sqrt{1- \cos^2 \varphi}}{\cos \varphi} \right\}
 \nonumber  \\ 
&&+2 \sqrt{c} \left\{\arcsin\left[\cos \varphi_n \right] 
-\arcsin\left[\cos \varphi \right]
\right\} \,,
\end{eqnarray}
where $\varphi_n= n \sqrt{c}/2 + \varphi$. According to Eq. (\ref{m(x)}) one obtains
\begin{eqnarray}
\label{cos_m_n_m_0}
&&m_0=\frac{2^{4/5}b^2}{c^2} \cos^4 \varphi \\
&&m_n=\frac{2^{4/5}b^2}{c^2} \cos^4 \varphi_n \, , 
\end{eqnarray}
which allows to express all trigonometric functions in Eq. (\ref{E_tot1}) 
in terms of $m_0$ and $m_n$, yielding
\begin{equation} 
\label{E_tot2}
E_{tot} = 2 \left\{ \sqrt{ \frac{2^{2/5}b}{m_n^{1/2}} -c} \, \, -
\, \sqrt{ \frac{2^{2/5}b}{m_0^{1/2}} -c} \, \, \right\} -\frac{cn}{2} \, , 
\end{equation}
and finally, taking into account Eq. (\ref{c}) for $c$, we arrive at the 
relation for the total losses
\begin{eqnarray}
\label{E_tot3}
E_{tot}(n,b)  = &&  2^{6/5} \sqrt{b} 
\left\{ \sqrt{ \frac{1}{\sqrt{m_n}}- \frac{1}{\sqrt{m^*}}} \right.  \\
&& - \left. \sqrt{ \frac{1}{\sqrt{m_0}}- \frac{1}{\sqrt{m^*}}} \, \, \, 
-\frac{n \sqrt{b}}{2^{9/5} \sqrt{m^*}} \right\} \, .\nonumber
\end{eqnarray}
Using   the approximation for the maximal mass
\begin{equation}
m^* \approx 
\left( n \sqrt{b}/p + q/p \right)^4 \, ,   
\end{equation}
which follows  from Eq. (\ref{m_n3}),  one obtains an explicit 
{\em approximate} relation  for the total losses and, thus, for the 
final velocity
\begin{equation} 
\label{v_fin}
v_n^{\prime \, 2}= \frac{m_0 v_0^2}{m_n} -\frac{2}{m_n} E_{tot}(n,b)
\end{equation}
in terms of of the chain length and the dissipation constant $b$.
Unfortunately, due to the fact that chains with optimal lengths
obviously do not have a maximum in their mass distribution, one cannot
use the previous relations to estimate the optimal chain length for a
given dissipation constant $b$, since these relations hold true only
for chains which do have a maximum.

Note that since the maximal mass $m^*$ depends only on the product 
$n\sqrt{b}$, the expression in curled brackets in the right-hand side 
of Eq. (\ref{E_tot3}) also depends only on this combination.  This suggests 
the following scaling relations for the final velocity for the chains 
with fixed $n\sqrt{b}$:
\begin{eqnarray}
\label{v_fin_scal}
&&v_n^{\prime \, 2}= m_n^{-1} -d \sqrt{b} \\
&&v_n^{\prime \, 2}= m_n^{-1} -d^{\prime}/ n \nonumber       \,,
\end{eqnarray}
where we take into account that $m_0=1$, $v_0=1$, and where
$d$ and $d^{\prime}$ are some constants which are defined by the 
particular value of $n\sqrt{b}$. 

\section{Conclusion}

We investigated analytically and numerically the transmission of
kinetic energy through one-dimensional chains of inelastically
colliding spheres, where the first and the last mass is fixed. For
the case of a constant coefficient of restitution we found that in the
chain with optimal energy transmission the mass of each particle is
given by the geometric average of its neighbors, i.e. the distribution
of the masses of the spheres is a monotonous, exponentially decreasing
function. This function is independent on the coefficient of
restitution $\epsilon$ where the special case of elastically colliding
particles ($\epsilon=1$) is included. We derived an expression for the
chain length $n^*$ which leads for a given $\epsilon$ to the optimal
energy transfer (provided the masses in between the first and last
mass have been chosen properly).

The situation changes qualitatively if we assume that the chain
consists of viscoelastic spheres for which the coefficient of
restitution depends on the impact velocity. Here, the optimal mass
distribution which leads to maximum energy transfer is not 
necessarily a
monotonous function. Depending on the chain length $n$ and on the
material parameters of the spheres it may reveal a pronounced
maximum. The part of the kinetic energy of the first particle, which
has not been transfered to the last one, we consider as losses of
energy.  These losses have been characterized as losses according to
incomplete transfer of momentum due to mass mismatch of the particles
(inertia losses) and losses due to the dissipative nature of particle
collisions (viscous losses). We develop a theory which describes the 
total energy losses along the chain, so that the optimal mass distribution,
minimizing the losses, may be obtained as a solution of a variational
equation.  We find a general solution to this 
nonlinear second-order differential equation. Implication 
of the boundary conditions yields, however, a transcendental equation, which 
one needs to solve numerically (in practice, we 
solve numerically the initial differential equation).  We observed that 
our variational theory agrees well with the results of the numerical 
optimization for the mass distribution, provided the dissipative material 
parameter is not too large. We also performed a direct verification of 
the basic approximation used in our variational approach.  

From the exact solution of the variational equation we obtained 
an analytical expression which relates the heaviest 
mass in the mass distribution to  the chain length and the dissipation 
constant. We found that this analytical expression, having no fitting 
parameters, is in good agreement with the numerical data. Using the 
exact solution for the optimal mass distribution we also found an expression 
for the total energy losses. This allowed to obtain scaling relations 
which show how the velocity of the last particle in the chain scales 
with the length of the chain $n$ and with the dissipation constant $b$, 
for the chain with the value of $n \sqrt{b}$ fixed.

It has been demonstrated before that for the case of
``thermodynamically-large'' granular systems the impact-velocity
dependence of the restitution coefficient, as it is given for
viscoelastic particles, may lead to qualitatively different behavior
as compared to systems with a constant restitution coefficient,
e.g.~\cite{Luding,NBTP,Spahn}.  The system investigated here may serve
as an example of the major influence of the velocity dependence of the
restitution coefficient even for relatively small (``lab-scale'') and
simple systems. Therefore, in general, the assumption of a constant
coefficient of restitution is an approximation which justification
cannot be assumed {\em \'a priori} but has to be checked for each
particular application.

\acknowledgments 
The authors thank Thomas Schwager for valuable discussion. We are 
also grateful for the referee for his/her  very helpful comments.

\end{multicols}                           
\vfill\newpage


\begin{references}
\bibitem{remarkmult} 
 In what follows we will consider the optimal mass distribution for the 
 case when the energy is transmitted from the first to the last particle
 only by subsequent from-left-to-right collisions, i.e. we will not analyse 
 the case when energy is transmitted by  multiple collisions between 
 subsequent particles. It may be shown that condition for such transmition
 of energy may be always satisfied by choosing appropriate distances 
 (depending on initial velocity $v_0$) between the 
 successive particles in the chain.
\bibitem{bdef} 
If we require the density $\rho$ (and not radii) to be constant for all spheres
Eq. (\ref{epsviab}) reads for particles $i$ and $j$ \begin{equation}
\epsilon=1-2^{-\frac15} b\! \left(\!\frac{4}{\pi\rho}\right)^\frac1{15}\!\! 
\left(\frac{m_i^\frac13
m_j^\frac13}{m_i^\frac13+m_j^\frac13}\right)^\frac15\!\!
\left(\frac{m_i+m_j}{m_i m_j}\right)^\frac15 v_{ij}^\frac15\,.\nonumber
\end{equation} Numerical calculations show that the assumption of
equal particle density does not change the results qualitatively,
however, the analysis will be much more complicated if not impossible.
\end{references}
\end{document}